\documentclass[conference]{IEEEtran}
\IEEEoverridecommandlockouts
\usepackage{cite}
\usepackage{amsmath,amssymb,amsfonts}
\usepackage[numbers]{natbib}
\usepackage{algorithmic}
\usepackage{graphicx}
\usepackage{textcomp}
\usepackage{xcolor}
\def\BibTeX{{\rm B\kern-.05em{\sc i\kern-.025em b}\kern-.08em
    T\kern-.1667em\lower.7ex\hbox{E}\kern-.125emX}}
\usepackage{xspace}
\usepackage{amsthm}
\usepackage{amsmath}
\usepackage{enumitem}
\usepackage{booktabs}
\usepackage{multirow}
\usepackage{graphicx}
\usepackage[normalem]{ulem}
\useunder{\uline}{\ul}{}
\usepackage{float}
\usepackage{caption}
\usepackage{subcaption}
\usepackage{balance}
\newcommand{\modelname}{\textsf{MT4SR}\xspace}
\usepackage{pifont}
\newcommand{\cmark}{\ding{51}}%
\newcommand{\xmark}{\ding{55}}%
\usepackage{blindtext}
\usepackage{hyperref}

\IEEEoverridecommandlockouts
\IEEEpubid{\makebox[\columnwidth]{978-1-6654-8045-1/22/\$31.00~\copyright2022 IEEE \hfill} \hspace{\columnsep}\makebox[\columnwidth]{ }}

\begin{document}

\title{Sequential Recommendation with Auxiliary Item Relationships via Multi-Relational Transformer}

\author{\IEEEauthorblockN{Ziwei Fan\IEEEauthorrefmark{1}, Zhiwei Liu\IEEEauthorrefmark{2}, Chen Wang\IEEEauthorrefmark{1}, Peijie Huang\IEEEauthorrefmark{3}, Hao Peng\IEEEauthorrefmark{4}, Philip S. Yu\IEEEauthorrefmark{1}}
\IEEEauthorblockA{\IEEEauthorrefmark{1}Department of Computer Science, University of Illinois Chicago, Chicago, USA \\
\{zfan20, cwang266, psyu\}@uic.edu}
\IEEEauthorblockA{\IEEEauthorrefmark{2}Salesforce AI Research, Palo Alto, USA; zhiweiliu@salesforce.com}
\IEEEauthorblockA{\IEEEauthorrefmark{3}South China Agricultural University, Guangzhou, China; pjhuang@scau.edu.cn}
\IEEEauthorblockA{\IEEEauthorrefmark{4}School of Cyber Science and Technology, Beihang University, Beijing, China; penghao@act.buaa.edu.cn}
}
\IEEEpubidadjcol
\maketitle

\begin{abstract}
Sequential Recommendation~(SR) models user dynamics and predicts the next preferred items based on the user history. Existing SR methods model the `was interacted before' item-item transitions observed in sequences, which can be viewed as an item relationship. However, there are multiple auxiliary item relationships, \textit{e.g.,} items from similar brands and with similar contents in real-world scenarios. Auxiliary item relationships describe item-item affinities in multiple different semantics and alleviate the long-lasting cold start problem in the recommendation. However, it remains a significant challenge to model auxiliary item relationships in SR.

To simultaneously model high-order item-item transitions in sequences and auxiliary item relationships, we propose a \textbf{M}ulti-relational \textbf{T}ransformer capable of modeling auxiliary item relationships for SR~(\modelname). Specifically, we propose a novel self-attention module, which incorporates arbitrary item relationships and weights item relationships accordingly. Second, we regularize intra-sequence item relationships with a novel regularization module to supervise attentions computations. Third, for inter-sequence item relationship pairs, we introduce a novel inter-sequence related items modeling module. Finally, we conduct experiments on four benchmark datasets and demonstrate the effectiveness of \modelname over state-of-the-art methods and the improvements on the cold start problem. The code is available in \url{https://github.com/zfan20/MT4SR}.
\end{abstract}

\begin{IEEEkeywords}
Sequential Recommendation, Self-Attention, Item Relationships
\end{IEEEkeywords}

\section{Introduction}
Sequential Recommendation~(SR) draws increasing attention due to its superior dynamic user modeling and scalability. SR models the dynamics in the sequence and predicts the next preferred item. 
SR learns dynamic user interests by modeling item-item transitions observed in sequences. These item-item transitions can be treated as a type of relationship between items with the temporal order, which we can define as `was interacted before.' Among existing SR advancements, including Markov Chain methods~\cite{rendle2010factorizing} and RNN-based methods~\cite{ma2019hierarchical}, Transformer architecture~\cite{vaswani2017attention} achieves great success and inspires many contributions because of the capability of modeling high-order item-item transitions. Several Transformer-based methods~\cite{kang2018self, sun2019bert4rec, fan2022sequential} demonstrate the effectiveness for SR. 

\begin{figure}
         \centering
         \includegraphics[width=0.48\textwidth]{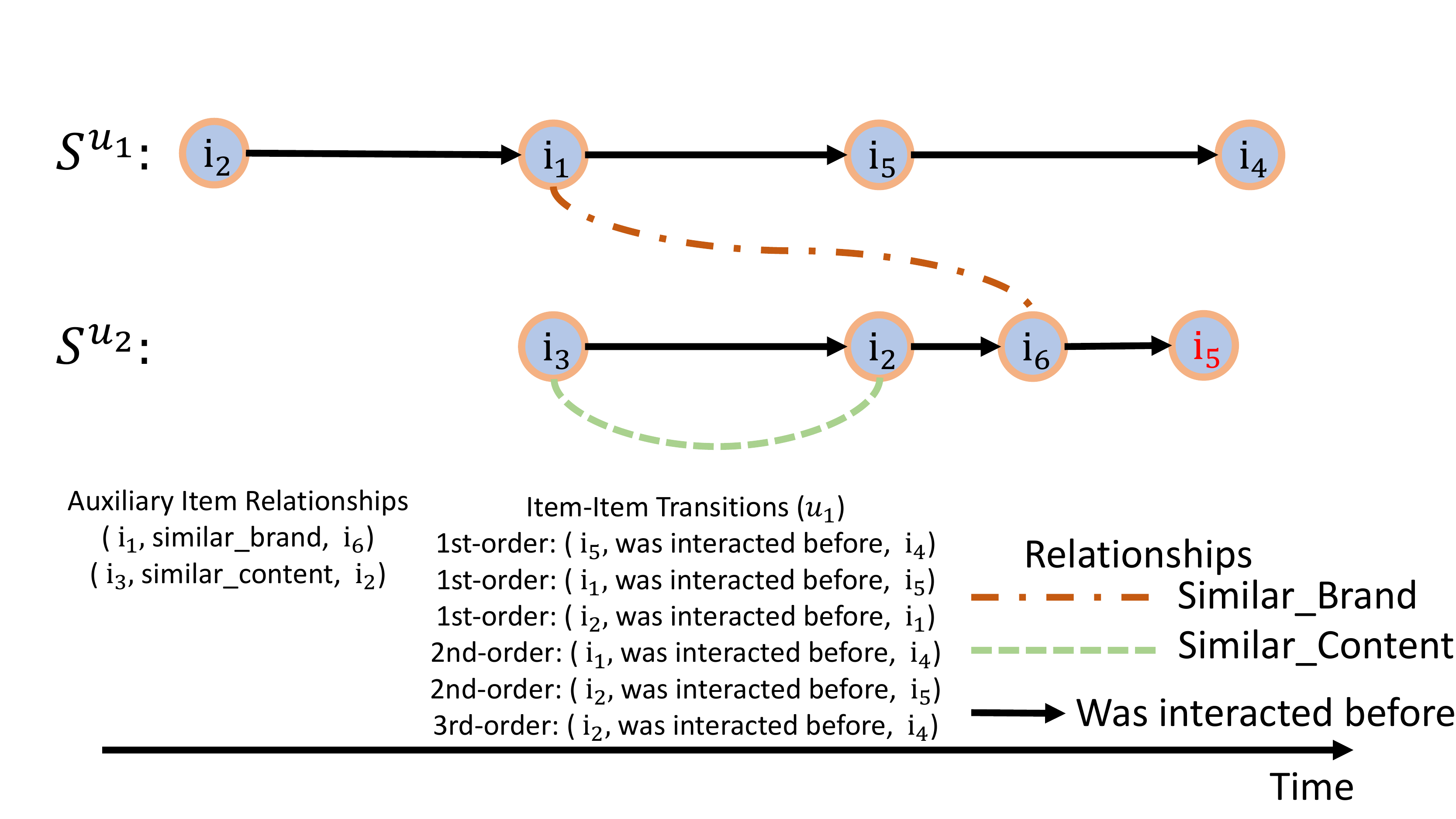}
         \caption{A toy example of two sequences with two item relationships `similar brand' and `similar content'. For item-item transitions from sequences~(black arrow), we define them as `was interacted before' asymmetric relationship. The auxiliary item relationships include intra-sequence related item pairs, \textit{e.g.,} ($i_3$, `similar\_content', $i_2$) can be observed as ($i_3$, was interacted before, $i_2$) in $\mathcal{S}^{u_2}$, and inter-sequence related item pairs, \textit{e.g.,} ($i_1$, `similar\_brand', $i_6$) crossing $\mathcal{S}^{u_1}$ and $\mathcal{S}^{u_2}$.}
         \label{fig:motivation}
\end{figure}

Modeling item-item transitions in SR is insufficient for satisfactory item embeddings learning because of the cold start problem. In real-world applications, there are multiple auxiliary item relationships, such as related items based on textual descriptions, search data, brands, and categorical connections. Auxiliary item relationships are a set of related item pairs under multiple relationships. It has been demonstrated that these auxiliary related item pairs benefit recommendation with great performance gains~\cite{kang2018recommendation, wang2021learning, wang2019kgat, 10.1145/3485447.3512273}. Although Transformer-based SR methods demonstrated the effectiveness, these methods cannot model auxiliary item relationships. 
In SR, higher-order item-item transitions can help but with limited contributions, as shown in Table~\ref{tab:transitions_hr}. 
In Table~\ref{tab:transitions_hr}, higher-order item transitions still can accurately predict testing item pairs.
However, 2nd-order and 3rd-order performances are worse than the 1st-order, which justifies the idea of Markov Chains models~\cite{rendle2010factorizing} and Transformer models~\cite{kang2018self, sun2019bert4rec, fan2022sequential}. From the bottom part of Table~\ref{tab:transitions_hr}, we can conclude that auxiliary item relationships have much higher hit ratios than pure item-item transitions and are potentially helpful for SR.

It is rather challenging to model auxiliary item relationships and high-order item transitions simultaneously. The challenges come from several perspectives: (1). the compatibility of item transitions and item relationships in self-attention; (2). proper supervision of related item pairs within sequences; (3). inter-sequences related item pairs are dominant.

First, the standard self-attention module~\cite{kang2018self, sun2019bert4rec, fan2022sequential, vaswani2017attention, liu2021augmenting, fan2021continuous, 9750359, peng2020m2, peng2022recursive} is not compatible to handle auxiliary item relationships as it only captures single item relationship observed in item-item transitions. 
Specifically, modeling auxiliary item relationships requires representing item relationships as attention values, which should be compatible with the scaled dot product self-attentions and demands theoretical support. Moreover, auxiliary item relationships are relationship-aware. Each relationship contributes to the next item recommendation unequally. For example, related item pairs observed from `co-searched' can better reflect users' intents than `similar in brand'.

Second, representing auxiliary item relationships via scaled dot product in self-attention still lacks correct supervisions and potentially misleads the attention calculation. For related item pairs observed in item-item transitions within sequences (intra-sequence), the dot product attention scores need to match well with relatedness signals, with the goal of correct guidance of proper attention computations. Without sufficient supervisions, attention scores from intra-sequence item relatedness are only random and free learnable parameters.

Third, most related item pairs are not intra-sequence but inter-sequences~\cite{qiu2020exploiting}, \textit{e.g.,} ($i_1$, $i_6$) in Fig.~\ref{fig:motivation}. However, inter-sequences related items enrich collaborative signals by connecting sequences with related items. The proportional ratio of intra-sequences related item pairs is small~($\leq 10\%$), as shown in Table~\ref{tab:seq_rel_ratio}. It indicates more than 90\% of related item pairs are inter-sequences. As intra-sequences related item pairs overlap with item-item transitions, these pairs only capture known information. Nevertheless, inter-sequences related item pairs significantly benefit the sequential recommendation. For example, in Fig.~\ref{fig:motivation}, given the history of $[i_3, i_2, i_6]$ of the user $u_2$, we fail to observe sufficient item-item transitions for signaling the next item $i_5$ because $S^{u_1}$ and $S^{u_2}$ have a small collaborative similarity based on histories. Moreover, $i_3$ and $i_6$ are cold items. However, with the help of inter-sequence related pair ($i_1$, `similar\_brand' $i_6$), we can draw additional collaborative connections between $u_1$ and $u_2$, and correctly recommend $i_5$. These additive connections incorporate more general collaborative signals rather than collaborative similarities based on interacted items.

\begin{table}[]
\centering
\caption{Testing item pair (next\_to\_last\_item, last\_item) (\textit{e.g.,} the ($i_5$, $i_4$) pair of user $u_1$ in Fig.~\ref{fig:motivation}.  Hit Ratio (HR) measures the percentage of testing item pairs captured by different orders of item transition pairs in training sequences and different item relationships (\textit{e.g.,} the ($i_1$, `similar\_brand', $i_6$) pair in Fig.~\ref{fig:motivation}). We adopt relationship `also viewed,' `also bought,' `bought together, and `buy after viewing' as auxiliary item relationships in four categories of the Amazon dataset.}
\label{tab:transitions_hr}
\begin{tabular}{@{}ccccc@{}}
\toprule
Dataset & Beauty & Toys & Tools & Office \\ \midrule
1st-order transition HR & 8.60\% & 8.06\% & 4.64\% & 11.74\% \\
2nd-order transition HR & 5.82\% & 4.12\% & 2.48\% & 8.95\% \\
3rd-order transition HR & 4.11\% & 2.22\% & 1.63\% & 6.44\% \\ \midrule
Total transition HR & 18.54\% & 14.41\% & 8.76\% & 27.14\% \\ \midrule
related item pairs HR & 22.94\% & 27.26\% & 16.93\% & 29.19\% \\ \bottomrule
\end{tabular}
\end{table}

\begin{table}[]
\centering
\caption{Intra-Sequence Related Item Pairs Coverage, which is calculated as $\left(\frac{1}{|\mathcal{U}|}\sum_{u\in\mathcal{U}}\frac{|\mathcal{I}\cap\{(v_i, v_j)\in\mathcal{S}^u\times \mathcal{S}^u\}|}{|\mathcal{S}^u| * |\mathcal{S}^u|}\right)$, where $\mathcal{I}$ refers to the set of related item pairs, $\times$ denotes set outer product, $\cap$ denotes the set intersection, $|\cdot|$ refers to size of set. The definitions of used symbols can be found in Section~\ref{subsec:prob_def}.}
\label{tab:seq_rel_ratio}
\begin{tabular}{@{}ccccc@{}}
\toprule
Dataset & Beauty & Toys & Tools & Office \\ \midrule
Sparsity & 4.58\% & 7.03\% & 3.37\% & 3.16\% \\ \bottomrule
\end{tabular}
\end{table}

In this paper, we develop a \textbf{M}ulti-relational \textbf{T}ransformer capable of processing auxiliary item relationships for SR~(\modelname). \modelname includes three core modules: (1). a multi-relational self-attention module designed for seamlessly incorporating auxiliary item relationships into the self-attention module; (2). a novel intra-sequence regularization term that supervises the related item pairs self-attention scores learning; (3). an explorative inter-sequences related items regularization that models related item pairs unobserved in sequences and further introduces additional collaborative signals for connecting similar behaviors. The contributions of this work are as follows:
\begin{itemize}[leftmargin=*]
    \item We propose a novel and general multi-relational self-attention Transformer framework to seamlessly incorporate auxiliary item relationships in SR.
    \item Inspired by the connection between self-attention and knowledge embeddings, we incorporate a novel item relatedness scoring in self-attention.
    \item We introduce two novel regularization terms for supervising intra-sequence related item pairs in the multi-relational self-attention and also explore inter-sequences related item pairs to explore additional collaborative signals across sequences.
    \item We demonstrate that \modelname outperforms state-of-the-art recommendation methods with improvements from 3.56\% to 21.87\% in all metrics on four benchmark datasets, including static methods, sequential methods, and methods using item relationship information.
\end{itemize}

\begin{table}[]
\centering
\caption{Model comparison. `H-R': Models auxiliary item relationships? `H-O': Models high-order information?}
\label{tab:related_comparison}
\begin{tabular}{@{}ccccc@{}}
\toprule
Capability & Personalized & Sequential & H-O & H-R \\ \midrule
BPRMF~\cite{rendle2010factorizing} & \cmark & \xmark & \xmark & \xmark \\
LightGCN~\cite{he2020lightgcn} & \cmark & \xmark & \xmark & \cmark \\
SASRec~\cite{kang2018self} & \cmark & \cmark & \cmark & \xmark \\
KGAT~\cite{wang2019kgat} & \cmark & \xmark & \cmark & \cmark \\
KGIN~\cite{wang2021learning} & \cmark & \xmark & \cmark & \cmark \\
RCF~\cite{xin2019relational} & \cmark & \xmark & \cmark & \cmark \\
MoHR~\cite{kang2018recommendation} & \cmark & \cmark & \xmark & \xmark \\
\modelname~(proposed) & \cmark & \cmark & \cmark & \cmark \\ \bottomrule
\end{tabular}
\end{table}

\section{Related Work}
\label{sec:related}
This section discusses existing methods related to our problem and the proposed method. We first introduce relevant methods in the sequential recommendation as it matches our task. Then we discuss existing methods for incorporating item relationships. Finally, we also introduce related works from knowledge graph recommendations because these works also model additional item knowledge. 
The summary and capabilities comparison of models are presented in Table~\ref{tab:related_comparison}.
\subsection{Sequential Recommendation}
Sequential Recommendation~(SR) predicts the next preferred item by modeling the chronologically sorted sequence of users' historical interactions. With the sequential modeling in the user's interaction sequence, SR captures the dynamic preference, which is latent in item-item transitions. One line of earliest works originates from the idea of Markov Chains, which are capable of learning item-item transition probabilities, including FPMC~\cite{rendle2010factorizing}. FPMC~\cite{rendle2010factorizing} captures only the first-order item transitions with low model complexity, assuming the next preferred item is only correlated to the previous interacted item. 
Fossil~\cite{he2016fusing} extends FPMC to learn higher-order item transitions and demonstrates the necessity of high-order item transitions in SR. 

The successful demonstration of sequential modeling from deep learning inspires research potentials of sequential models for SR, including Recurrent Neural Network~(RNN)~\cite{quadrana2017personalizing, ma2019hierarchical, zheng2019gated}, Convolution Neural Network~(CNN)~\cite{tang2018personalized, ma2019hierarchical}, and Transformer~\cite{kang2018self, sun2019bert4rec, fan2022sequential}. The representative work of RNN for SR is GRU4Rec~\cite{hidasi2015session}, which adopts the Gated Recurrent Unit~(GRU) in the session-based recommendation. Another line of SR is CNN-based methods, such as Caser~\cite{tang2018personalized}. Caser~\cite{tang2018personalized} treats the interaction sequence with item embeddings as an image and applies convolution operators to learn local sub-sequence structures. The recent success of self-attention-based Transformer~\cite{vaswani2017attention} architecture provides more possibilities in SR due to its capability of modeling all pair-wise relationships within the sequence, which is the limitation of RNN-based methods and CNN methods. SASRec~\cite{kang2018self} is the first work adopting the Transformer for SR and demonstrates its superiority. BERT4Rec~\cite{sun2019bert4rec} extends the SASRec to model bi-directional relationships in Transformers, with the inspiration of BERT~\cite{devlin2018bert}. 
TiSASRec~\cite{li2020time} further incorporates time difference information in SASRec. 
FISSA~\cite{lin2020fissa} explores latent item similarities in SR.
DT4Rec~\cite{fan2021modeling} and STOSA~\cite{fan2022sequential} model items as distributions instead of vector embedding and are state-of-the-art SR methods with implicit feedback.

Despite the recent success of SR methods, they still fail to incorporate heterogeneous item relationships into the modeling of item-item transitions, especially in high-order transitions. Distinctly, the proposed \modelname can model both item-item purchase transitions and additional item relationships in a unified framework, which can be easily extended to various numbers of relationships.

\subsection{Item Relationships-aware Recommendation}
Some methods propose to utilize extra item relationships~\cite{liu2020basket,DBLP:journals/tois/PengZDYZY22} to enhance the representation capability of item embeddings. For example, Chorus~\cite{wang2020make} specifically models substitute and complementary relationships between items in the continuous-time dynamic recommendation scenario. RCF~\cite{xin2019relational} proposes to model item relationships in a two-level hierarchy in a graph learning framework. UGRec~\cite{zhao2021ugrec} extends the idea of RCF and adopts the translation knowledge embedding approach within the graph recommendation framework to model both directed and undirected relationships for the recommendation. MoHR~\cite{kang2018recommendation} is the most relevant work to this paper. MoHR incorporates item relationships into first-order user-item translation scoring and proposes optimizing the next relationship prediction, which can identify the importance of each relationship in the dynamic sequence.

Although these methods significantly improve the recommendation, they still obtain sub-optimal performance in recommendation and efficiency. Chorus can only handle substitute and complementary relationships for sequential recommendation while more item relationships exist, and identifying the significance of relationships is also crucial. RCF and UGRec both rely on the graph modeling framework, which sometimes requires a large amount of graphical memory due to the exponential growth neighbors. Furthermore, neither RCF nor UGRec can handle dynamic user preferences. MoHR only models the first-order translation between user and item under the relationship space~\cite{kang2018self}.

\subsection{Knowledge Graph Recommendation}
Knowledge graph recommendation~\cite{huang2018improving, wang2019kgat, wang2021learning, liu2022federated, wang2021dskreg, yang2022large} originates from knowledge embeddings learning, where the knowledge graph consists of the triplets describing entities and their relationships. The classical line of knowledge graph recommendation is embedding-based methods, which adopt knowledge embedding techniques to learn entity and relation embeddings, such as TransE~\cite{bordes2013translating}, and DistMult~\cite{yang2014embedding}. The representative work is CKE~\cite{zhang2016collaborative}. CKE utilizes TransE to learn knowledge embeddings and regularizes the matrix factorization. KTUP applies TransE to model both knowledge triplets and user-item interactions. Another line of work is path-based methods, in which RippleNet~\cite{wang2018ripplenet} is the representative work. RippleNet starts paths from each user and aggregates item embeddings with the path. The most state-of-the-art methods are based on collaborative knowledge graphs, including KGAT~\cite{wang2019kgat} and KGIN~\cite{wang2021learning}. Both KGAT and KGIN combine the item knowledge graph and the user-item interaction graph as a unified graph. KGAT applies TransE scores as attention weights for node message aggregation. KGIN extends KGAT by modeling paths as intents. 

\section{Preliminaries}

\subsection{Problem Definition}\label{subsec:prob_def}
Given a set of users $\mathcal{U}$ and items $\mathcal{V}$, and the associated interactions, 
we first sort the interacted items of each user $u\in\mathcal{U}$ chronologically in a sequence as $\mathcal{S}^{u}={[v^{u}_1,v^u_2,\dots,v^u_{|\mathcal{S}^{u}|}]}$, where $v^{u}_i\in\mathcal{V}$ denotes the $i$-th interacted item in the sequence. In addition to the interaction sequence, there are item relationship pairs $\{(v_i, r, v_j)\in\mathcal{I}\}$ with a number of relationships $\{r\in\mathcal{R}\}$, where $\{v_i\in\mathcal{V}\}$ and $\{v_j\in\mathcal{V}\}$. $\mathcal{I}_{v, r}$ refers to the set of items related to the item $v$ by the relationship $r$. The goal of SR is to recommend a top-N ranking list of items as the potential next items in a sequence. Formally, we should predict $p\left( v_{|\mathcal{S}^{u}|+1}^{(u)}=v \left|  \mathcal{S}^{u},\mathcal{I} \right.\right)$.

\subsection{Self-Attention for SR}
We build the proposed model upon the original self-attention module as the sequence encoder in this paper, and we first introduce it before presenting our model.
To be specific,
given a user's action sequence $\mathcal{S}^u$ and the predefined maximum sequence length $L$, the sequence is truncated by removing earliest items if $|\mathcal{S}^u|>L$ or padded with zeros to obtain a fixed length sequence $s=(s_1, s_2, \dots, s_L)$. An item embedding matrix $\mathbf{M}\in \mathbb{R}^{|\mathcal{V}|\times d}$ is defined, where $d$ is the latent dimension size. 
A trainable
positional embedding $\mathbf{P}\in \mathbb{R}^{L\times d}$ is added with item embeddings within the sequence to get the sequence embedding:
\begin{equation}
    \label{eq:seq_emb}
    \mathbf{E}_{\mathcal{S}^{u}} = [\mathbf{m}_{s_1}+\mathbf{p}_{s_1}, \mathbf{m}_{s_2}+\mathbf{p}_{s_2}, \dots, \mathbf{m}_{s_n}+\mathbf{p}_{s_L}].
\end{equation}
Specifically, self-attention~(SA) adopts scaled dot-products between item embeddings in the sequence to obtain their pair-wise correlations, which are as follows:
\begin{equation}
\label{eq:sa}
    \text{SA}(\mathbf{E}_{\mathcal{S}^{u}}) = \text{softmax}\left(\frac{\mathbf{Q}\mathbf{K}^\top}{\sqrt{d}}\right) \mathbf{V}, 
\end{equation}
where $\mathbf{Q}=\mathbf{E}_{\mathcal{S}^{u}}\mathbf{W}_Q$, $\mathbf{K}=\mathbf{E}_{\mathcal{S}^{u}}\mathbf{W}_K$, and $\mathbf{V}=\mathbf{E}_{\mathcal{S}^{u}}\mathbf{W}_V$. $\mathbf{W}_Q\in\mathbb{R}^{d\times d}$, $\mathbf{W}_K\in\mathbb{R}^{d\times d}$, and $\mathbf{W}_V\in\mathbb{R}^{d\times d}$ are learnable weight matrices for key, query, and value transformations.
Other components in Transformer are utilized in SASRec, including the point-wise feed-forward network, residual connection, and layer normalization.

\begin{figure*}
         \centering
         \includegraphics[width=0.8\textwidth]{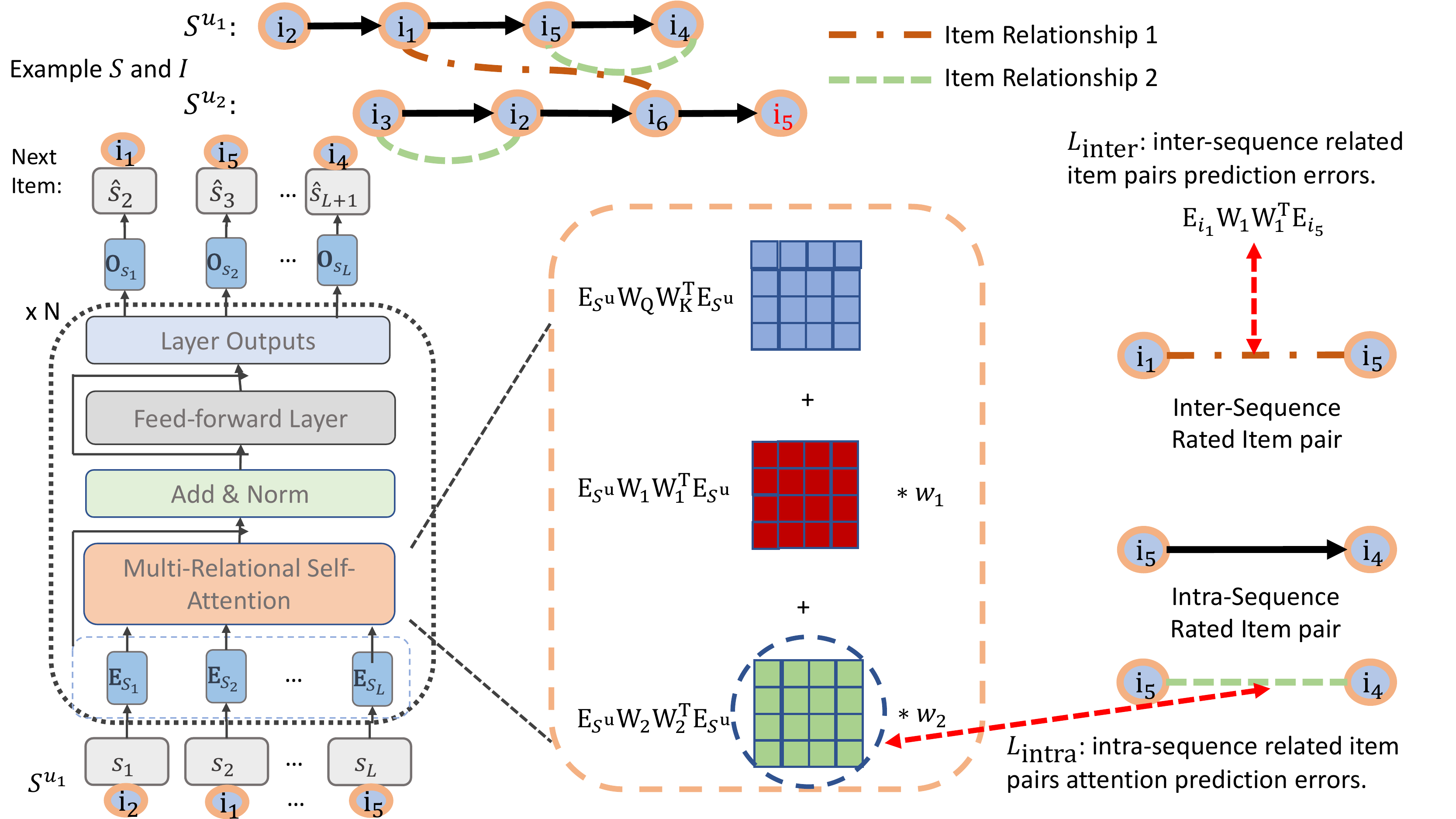}
         \caption{Model Architecture of \modelname. Note that intra-sequence and inter-sequences related item pairs can appear in all relationship. }
         \label{fig:model_archi}
\end{figure*}
\section{Proposed Model}
This section introduces the proposed multi-relational self-attention for SR, \modelname, which consists of three components. Figure~\ref{fig:model_archi} shows the overall model architecture of \modelname. The first component is the multi-relational self-attention module. 
The second component is the intra-sequence item relationships modeling for fitting the related item pairs observed in the sequence. The last module is inter-sequences related items modeling, exploring item pairs outside sequences.

\subsection{Self-Attention with Auxiliary Item Relationships}
The existing self-attention modules~\cite{vaswani2017attention, devlin2018bert} typically only handle a single item relationship in the sequence, which is `was interacted before' in SR. A relevant work MoHR~\cite{kang2018recommendation} can process additional related items with various relationships, but it can only handle first-order item transitions. Self-attention models all item-item pairs within the sequence and naturally considers high-order item transitions. There remain challenges in modeling sequential dynamics with auxiliary related item pairs and high-order transitions simultaneously. Different from item-item transitions, modeling auxiliary relationships needs to be relationship-aware. To address both challenges, we introduce the Multi-Relational Self-Attention~(MRSA) to incorporate relationship types information into the attention weight calculation. We first discuss the connection between existing dot-product attention and knowledge embeddings and conclude that the scaled dot-product can be interpreted as a variant of knowledge embeddings. Based on this connection, we introduce auxiliary item relationships modeling for enhancing self-attention.

\subsubsection{Connection between Self-Attention and Knowledge Embeddings}
We first discuss the connections and the differences of existing dot-product attention and the knowledge embedding DistMult~\cite{yang2014embedding}. From Eq.~(\ref{eq:sa}), we extract the dot product component in self-attention calculation for a specific item pair $(v_{s_i}, v_{s_j})$, which is as follows:
\begin{equation}
\begin{aligned}
\label{eq:sa_onepair}
    \text{Att}(v_{s_i}, v_{s_j}) = \mathbf{Q}_{v_{s_i}}\mathbf{K}_{v_{s_j}}^\top &= \mathbf{E}_{v_{s_i}}\mathbf{W}_Q\mathbf{W}_K^\top\mathbf{E}_{v_{s_j}}^\top\\
    &=\mathbf{E}_{v_{s_i}}\mathbf{W}_{QK}\mathbf{E}_{v_{s_j}}^\top
\end{aligned}
\end{equation}
where $\mathbf{E}_{v_{s_i}}\in \mathbb{R}^{1\times d}$ and $\mathbf{E}_{v_{s_j}}\in \mathbb{R}^{1\times d}$ denote the item embeddings of item $v_{s_i}$ and $v_{s_j}$ in $\mathcal{S}^u$ repspectively, $\mathbf{W}_Q\in \mathbb{R}^{d\times d}$, $\mathbf{W}_K\in \mathbb{R}^{d\times d}$ are weight matrices in self-attention, and $\mathbf{W}_{QK}=\mathbf{W}_Q\mathbf{W}_K^\top$. The attention calculation brings up the closeness between self-attention and knowledge embedding scoring functions, including ANALOGY~\cite{liu2017analogical} and DistMult~\cite{yang2014embedding}. Specifically, given a knowledge triplet $(h, r, t)$, the scoring function of DistMult is defined as follows~\cite{ji2021survey}:
\begin{equation}
\label{eq:distmult}
    f_r(h, t) = h\cdot\text{diag}(\mathbf{w}_r)\cdot t^\top,
\end{equation}
where $h$ and $t$ are head and tail entity embeddings, $\mathbf{w}_r\in\mathbb{R}^{d}$ is the relation weight embedding of relation $r$, and the $\text{diag}(\mathbf{w}_r)\in\mathbb{R}^{d\times d}$. The scoring function of ANALOGY is:
\begin{equation}
\label{eq:analogy}
    f_r(h, t) = h\cdot\mathbf{W}_r\cdot t^\top,
\end{equation}
where $\mathbf{W}_r\in\mathbb{R}^{d\times d}$ is a normal relation matrix that $\mathbf{W}_r\mathbf{W}_r^\top=\mathbf{W}_r^\top\mathbf{W}_r$. 

We can observe the \textit{connection} among Eq.~(\ref{eq:sa_onepair}), Eq.~(\ref{eq:distmult}), and Eq.~(\ref{eq:analogy}) if we view the $\mathbf{W}_{QK}$ as the relation weight matrix of the relationship `was interacted before'. To this end, we can conclude that the dot-product attention defined in the self-attention module can be viewed as a variant of the knowledge embedding scoring function. 

However, there is a significant \textit{difference} among them. $\mathbf{W}_{QK}$, as a relationship weight matrix, is not a normal matrix, which indicates that the relationship modeled in the dot-product attention are asymmetric, even in the bi-directional version BERT~\cite{devlin2018bert}~(removing the causality masking in Transformer). This is reasonable in the SR next-item prediction task because the temporal order matters in the sequential modeling~\cite{kang2018self}. By comparing with DistMult and ANALOGY, DistMult encodes the relation as a vector, and ANALOGY constrains the weight matrix as a normal matrix, lacking sufficient representation flexibility or introducing optimization difficulty. 

\subsubsection{Multi-Relational Self-Attentions}
To enhance the dot-product self-attention module with auxiliary item relationships modeling, we build upon ANALOGY to calculate the item relatedness scoring $\text{MRSA}(\mathbf{E}_{\mathcal{S}^{u}})$ as follows:
\begin{equation}
\label{eq:rsa}
    \text{softmax}\left(\frac{\mathbf{Q}\mathbf{K}^\top+\sum\limits_{r\in\mathcal{R}}w_r \mathbf{E}_{\mathcal{S}^{u}}\mathbf{W}_r\mathbf{W}_r^\top \mathbf{E}_{\mathcal{S}^{u}}^\top}{\sqrt{d}}\right) \mathbf{V}, 
\end{equation}
where $\mathcal{R}$ denotes the set of relationships,  $\mathbf{W}_r \in\mathbb{R}^{d\times d}$ is the learnable weight matrix of relationship $r$, $w_r$ is a learnable scalar for controlling the weight of the relationship $r$. Note that $\mathbf{W}_r\mathbf{W}_r^\top$ is a normal matrix, similar to the definition in ANALOGY without constraints, indicating the capability of modeling auxiliary item relationships in SR. MRSA can handle the arbitrary number of item relationships and model these item pairs in high-order item transitions, as self-attention models all item pairs within the sequence. Note that the number of relationship $|\mathcal{R}|$ is small, \textit{e.g.,} $|\mathcal{R}|\leq 10$.

\subsection{Intra-Sequence Item Relationships Supervision}
Based on the calculation of MRSA defined in Eq.~(\ref{eq:rsa}), the auxiliary item relatedness scorings do not use the input related item pairs, \textit{i.e.,} $\mathcal{I}$, as the supervised signals to guide the computations for accurate attentions. Without the supervision of $\mathcal{I}$, the additive multi-relational attention component acts only as extra free parameters. To resolve this issue, we propose a regularization term that measures the errors between the predictions of intra-sequence related item pairs and the ones of ground truth related item pairs as follows:
\begin{equation}
\begin{aligned}
    \mathcal{L}_{\text{intra}} = -&\sum_{i=1}^{\mathcal{S}^{u}}\sum_{j>i}^{\mathcal{S}^u}\sum_{r\in\mathcal{R}}\biggl[\mathbf{I}\left((v_i, v_j)\in \mathcal{I}_r\right)*\log\sigma\left(f_r(v_i,v_j)\right)\\ 
    & + \left(1-\mathbf{I}\left((v_i, v_j)\in \mathcal{I}_r\right)\right) \log\sigma\left(1-f_r(v_i,v_j)\right)\biggr],
\end{aligned}
\end{equation}
where $\mathcal{I}_r$ refers to the set of item pairs with the relationship $r$, $\mathbf{I}\left((v_i, v_j)\in \mathcal{I}_r\right)$ is indicator function with value of 1 when $(v_i, v_j)$ exists in $\mathcal{I}_r$ and 0 otherwise, $\sigma(\cdot)$ is the sigmoid function, $f_r(v_i,v_j)=\mathbf{E}_{v_i}\mathbf{W}_r\mathbf{W}_r^\top \mathbf{E}_{v_j}^\top$ denotes the relatedness prediction score of item pair $(v_i, v_j)$ in relationship $r$, which is defined in Eq.~(\ref{eq:rsa}). $\mathcal{L}_{\text{intra}}$ measures the relatedness prediction errors of all intra-sequence item pairs. When $\mathcal{L}_{\text{intra}}$ is optimized to be close to 0, relatedness of all intra-sequence item pairs are correctly predicted, \textit{i.e.,} $\mathbf{E}_{\mathcal{S}^{u}}\mathbf{W}_r\mathbf{W}_r^\top \mathbf{E}_{\mathcal{S}^{u}}^\top$ in Eq.~(\ref{eq:rsa}) has correct attention computations.

\subsection{Inter-Sequences Related Items Modeling}
There are only limited portions of intra-sequence related item pairs ($<$10\% shown in Table~\ref{tab:seq_rel_ratio}). The inter-sequences item pairs help connect item transitions across sequences and incorporate more users' collaborative signals from connected sequences. 
To fully explore and utilize the inter-sequences signals, we propose a novel regularization term, which describes the inter-sequences item pairs and is defined as follows:
\begin{equation}
\begin{aligned}
    \mathcal{L}_{\text{inter}} = -\sum_{r\in\mathcal{R}}\sum_{v_i\in\mathcal{I}_r}& \biggl[\log\sigma\left(f_r(v_i,v_{j+})\right)\\
    & + \log\sigma\left(1-f_r(v_i,v_{j-})\right)\biggr],
\end{aligned}
\end{equation}
where $v_{j+}\in\mathcal{I}_{v_i, r}$ is a positive item with the relationship $r$ with the item $v_i$, $v_{j-}\in \mathcal{V}\setminus\mathcal{I}_{v_i, r}$ is a negative sampled item without relationship $r$ connection with the item $v_i$. The $\mathcal{L}_{\text{inter}}$ regularization term reinforces the relatedness between item pairs that are inter-sequences. The fundamental difference between $\mathcal{L}_{\text{intra}}$ and $\mathcal{L}_{\text{inter}}$ is that $\mathcal{L}_{\text{intra}}$ focuses only the item pairs within sequences while $\mathcal{L}_{\text{inter}}$ can explore item pairs that never exist in training sequences, \textit{i.e.,} inter-sequences. The $\mathcal{L}_{\text{intra}}$ and $\mathcal{L}_{\text{inter}}$ are complementary and benefits the exploration of additional sequential collaborative signals.

\subsection{Prediction Layer}
In the prediction layer, we still apply the point-wise feedforward networks~(FFN), residual connections, dropout, and layer normalization techniques for inferring the next item embedding. The detailed calculation can be found in related papers~\cite{vaswani2017attention, kang2018self}. To be specific, the overall process includes:
\begin{equation}
    \begin{aligned}
    \label{eq:pred}
        \text{F}_{\mathcal{S}^{u}} &= \text{FFN}(\text{LN}\left(\text{MRSA}(\mathbf{E}_{\mathcal{S}^{u}}))\right)\\
        \text{O}_{\mathcal{S}^{u}} &= \text{F}_{\mathcal{S}^{u}} + \text{Dropout}\left(\text{F}_{\mathcal{S}^{u}}\right),
    \end{aligned}
\end{equation}
where $\text{LN}$ denotes the layer normalization, the process in Eq.~(\ref{eq:pred}) can be stacked for multiple layers by feeding the output sequence embedding $\text{O}_{\mathcal{S}^{u}}$ to the next \modelname block. By having $K$ number of layers, we use the output sequence embeddings from the last layer $\text{O}_{\mathcal{S}^{u}}^K$ for generating the next item $v_i$ prediction score as follows:
\begin{equation}
    r(\mathcal{S}^{u}_L,v_i) = \text{O}_{L}^KE_{v_i}.
\end{equation}
$r(\mathcal{S}^{u}_L,v_i)$ indicates the possibility of item $v_i$ being the next item after the sequence $\mathcal{S}^{u}$ with the length of $L$. We calculate the $r(\mathcal{S}^{u}_L,v_i)$ over all candidate items $v_i$ to generate the ranked item list for top-N next item recommendation by sorting the scores in descending order.

\subsection{Loss}
The final loss consists of three components, the recommendation loss, $\mathcal{L}_{\text{intra}}$, and $\mathcal{L}_{\text{inter}}$. We adopt the cross-entropy loss to measure the next-item prediction error on each position in the sequence, which is defined as follow:
\begin{equation}
    \mathcal{L}_{\text{pred}} = -\sum_{\mathcal{S}^u\in\mathcal{S}}\sum_{t=1}^{|\mathcal{S}^u|}\left[\log \left(\sigma(r_{\mathcal{S}^{u}_t, j^{+}})\right) + \log \left(1-\sigma(r_{\mathcal{S}^{u}_t, j^{-}})\right)\right],
\end{equation}
where $j^{+}$ is the ground truth next item at step $t$ in $\mathcal{S}^u$, $j^{-}$ is sampled from the items that the user $u$ has no interaction with, and $\sigma(\cdot)$ denotes the sigmoid function. The final loss is defined as:
\begin{equation}
    \mathcal{L} = \mathcal{L}_{\text{pred}} + \alpha \mathcal{L}_{\text{intra}} + \beta \mathcal{L}_{\text{inter}}+\lambda||\Theta||_2^2,
\end{equation}
where $\Theta$ consists of all learnable parameters in \modelname, $\alpha$, $\beta$, and $\lambda$ are hyper-parameters.

\section{Experiments}
In this section, we demonstrate the effectiveness of \modelname in top-N recommendation results and detailed analysis. We answer the following research questions~(RQs):
\begin{itemize}[leftmargin=*]
    \item \textbf{RQ1}: Does \modelname provide better recommendations than existing methods?
    \item \textbf{RQ2}: How sensitive is the recommendation performance with varying $\alpha$ and $\beta$?
    \item \textbf{RQ3}: How does each proposed module affect the recommendation performance?
    \item \textbf{RQ4}: Where do improvements of \modelname come from?
\end{itemize}


\begin{table}[H]
\centering
\caption{Datasets Statistics After Preprocessing}
\label{tab:data_stat}
\begin{tabular}{@{}ccccc@{}}
\toprule
Dataset & Beauty & Toys & Tools & Office \\ \midrule
\#users & 22,363 & 19,412 & 16,638 & 4,905 \\
\#items & 12,101 & 11,924 & 10,217 & 2,420 \\
\#ratings & 198,502 & 167,597 & 134,476 & 53,258 \\
density & 0.05\% & 0.07\% & 0.08\% & 0.44\% \\ \midrule
avg ratings/user & 8.3 & 8.6 & 8.1 & 10.8 \\ \midrule
avg ratings/item & 16.4 & 14.0 & 13.1 & 22.0 \\ \midrule
\begin{tabular}[c]{@{}c@{}}\#related item\\ pairs\end{tabular} & 403,724 & 624,213 & 300,514 & 58,829 \\ \midrule
avg pairs/item & 33.4 & 52.3 & 29.4 & 24.3 \\ \bottomrule
\end{tabular}
\end{table}

\subsection{Datasets and Preprocessing}
We conduct the experiments on four benchmark datasets from Amazon review datasets across various domains. Amazon datasets are known for high sparsity and rich meta information of items. There are also several sub-categories of rating reviews in Amazon datasets. We select Beauty, Tools, Toys, and Office sub-categories because of the wide adoption in \cite{fan2021modeling, fan2022sequential, kang2018self, sun2019bert4rec}. Amazon datasets have four types of item relationships, including `also viewed,' `also bought,' `bought together, and `buy after viewing.' Following~\cite{sun2019bert4rec, kang2018self, he2017translation, fan2021modeling, fan2022sequential}, we treat the presence of ratings as positive interactions and also adopt the 5-core settings by filtering out users with less than 5 interactions. We use timestamps of ratings to sort interactions and form the sequence for each user. The last interacted item is used for testing, and the second last interacted item is used for validation. 
Details of datasets\footnote{\url{https://jmcauley.ucsd.edu/data/amazon/}} are shown in Table~\ref{tab:data_stat}.

\subsection{Evaluation}
We \textbf{rank all items} instead of the biased negative sampling evaluation~\cite{krichene2020sampled} for accurate models comparison. We adopt the standard top-N ranking evaluation metrics to evaluate the recommendation performance, including Recall@N, NDCG@N, and MRR. Recall@N measures the ratio of the ground truth positive item appearing in the top-N recommendation list. NDCG@N considers the ranking position of the positive item in the top-N list by assigning different weights in ranking positions. MRR evaluates the performance for the entire ranking list while also considering ranking positions. We report the averaged test metric results over all users based on the best validation performance. We report the performances when $N=5$ and $N=10$, which are also adopted by \cite{sun2019bert4rec, kang2018self, he2017translation, fan2021modeling, fan2022sequential}.

\subsection{Baselines}
We compare the proposed \modelname with the following baselines in three groups. The first group includes static recommendation methods, including BPR~\cite{rendle2012bpr} and LightGCN~\cite{he2020lightgcn}. The second group includes sequential recommendation methods: Caser~\cite{tang2018personalized}, SASRec~\cite{kang2018self}, BERT4Rec~\cite{sun2019bert4rec}, and STOSA~\cite{fan2022sequential}. The third group consists of recommendation methods with item relationships modeling, including knowledge graph recommendation methods KGAT~\cite{wang2019kgat} and KGIN~\cite{wang2021learning} as well as the sequential method MoHR~\cite{kang2018recommendation}. We also include the RCF~\cite{xin2019relational} as the baseline model in the third group. 
We \textbf{grid search} all parameters and report the test performance based on the best validation result. We search the embedding dimension in $\{64, 128\}$ for all baselines, max sequence length from $\{50, 100\}$, learning rate from $\{10^{-3},10^{-4}\}$, the L2 regularization weight from $\{10^{-1}, 10^{-2}, 10^{-3}\}$, dropout rate from $\{0.3, 0.5, 0.7\}$. For sequential methods, we search the number of layers from $\{1,2,3\}$ and the number of heads in $\{1,2,4\}$. We adopt the early stopping strategy that model optimization stops when the validation MRR does not increase for 50 epochs.
The details of hyperparameters searching and implementations are in Appendices. 
\begin{itemize}[leftmargin=*]
    \item \textbf{BPR:} BPR is the most classical collaborative filtering method for top-N recommendation of implicit feedbacks.
    \item \textbf{LightGCN:} LightGCN is the state-of-the-art static graph recommendation method, which considers high-order collaborative signals inherent in user-item graph. We search number of layers from $\{1,2,3\}$, and node dropout from $\{0.1, 0.3, 0.5, 0.7\}$.
    \item \textbf{Caser:} A CNN-based sequential recommendation method that applies convolution operators to the sequence embedding matrix, which can be viewed as an image. We search the length $L$ from $\{5, 10\}$, and $T$ from $\{1, 3, 5\}$.
    \item \textbf{SASRec:} The state-of-the-art sequential method that builds upon the Transformer. We search the dropout rate from $\{0.3, 0.5, 0.7\}$.
    \item \textbf{BERT4Rec:} This method extends SASRec to model bidirectional item transitions with Cloze objective. We search the mask probability from the range of $\{0.1, 0.2, 0.3, 0.5, 0.7\}$.
    \item \textbf{STOSA:} The most recent state-of-the-art sequential recommendation method with only modeling implicit feedbacks. It proposes a novel self-attention that models items as distributions and adopts Wasserstein distance as attentions. We search the $\lambda$ in STOSA from $[0, 1]$ with increment of 0.1.
    \item \textbf{KGAT:} KGAT is one of the state-of-the-art recommendation methods with modeling of item knowledge. It learns item embeddings by fitting the collaborative signals and item relationships in the knowledge perspective. We search the number of layers from $\{1, 2\}$, node dropout probability from $\{0.1, 0.5\}$, and knowledge graph regularization weight from $\{0.1, 1.0, 5.0\}$.
    \item \textbf{KGIN:} KGIN is the most recent item knowledge graph-based recommendation method. It extends the idea of KGAT and learns intents as multi-hops paths of the collaborative knowledge graph. We search the similarity regularization from $\{1e-4, 1e-5\}$, node dropout probability from $\{0.3, 0.4, 0.5\}$, message dropout probability from $\{0.1, 0.3\}$, and the number of hops from $\{1,2,3\}$.
    \item \textbf{MoHR:} MoHR is the closest work to \modelname. It models the item relationships in the sequential recommendation setting and also proposes the idea of next relationship prediction. For MoHR specific hyperparameters $\alpha$, $\beta$ and $\gamma$, we search the $\alpha$ from $\{0.1, 0.3, 0.5\}$, $\beta$ from $\{0.01, 0.05, 0.1\}$, and $\gamma$ from $\{0.01, 0.05, 0.1\}$.
\end{itemize}

\begin{table*}[]
\centering
\scriptsize
\caption{Overall Performance Comparison Table. The best and second-best results are bold and underlined, respectively. `Improve.' is the relative improvement against the second-best baseline performance.}
\label{tab:overall_perf}
\begin{tabular}{@{}cccccccccccccc@{}}
\toprule
Dataset & Metric & BPRMF & LightGCN & Caser & SASRec & BERT4Rec & RCF & STOSA & KGAT & KGIN & MoHR & \modelname & Improv. \\ \midrule
 & Recall@5 & 0.0300 & 0.0287 & 0.0309 & 0.0416 & 0.0396 & 0.0412 & 0.0504 & 0.0219 & 0.0319 & {\ul 0.0529} & \textbf{0.0579} & +9.29\% \\
 & NDCG@5 & 0.0189 & 0.0174 & 0.0214 & 0.0274 & 0.0257 & 0.0264 & {\ul 0.0351} & 0.0130 & 0.0200 & 0.0349 & \textbf{0.0390} & +11.22\% \\
Beauty & \multicolumn{1}{l}{Recall@10} & 0.0471 & 0.0468 & 0.0407 & 0.0633 & 0.0595 & 0.0601 & 0.0707 & 0.0373 & 0.0540 & {\ul 0.0829} & \textbf{0.0859} & +3.56\% \\
 & \multicolumn{1}{l}{NDCG@10} & 0.0245 & 0.0233 & 0.0246 & 0.0343 & 0.0321 & 0.0336 & 0.0416 & 0.0180 & 0.0271 & {\ul 0.0445} & \textbf{0.0480} & +7.75\% \\
 & MRR & 0.0216 & 0.0203 & 0.0231 & 0.0291 & 0.0294 & 0.0306 & 0.0360 & 0.0159 & 0.0230 & {\ul 0.0386} & \textbf{0.0408} & +5.61\% \\ \midrule
 & Recall@5 & 0.0216 & 0.0231 & 0.0129 & 0.0284 & 0.0189 & 0.0256 & 0.0312 & 0.0163 & 0.0221 & {\ul 0.0481} & \textbf{0.0536} & +11.50\% \\
 & NDCG@5 & 0.0139 & 0.0152 & 0.0091 & 0.0194 & 0.0123 & 0.0153 & 0.0217 & 0.0101 & 0.0142 & {\ul 0.0340} & \textbf{0.0379} & +11.70\% \\
Tools & \multicolumn{1}{l}{Recall@10} & 0.0334 & 0.0359 & 0.0193 & 0.0427 & 0.0319 & 0.0354 & 0.0468 & 0.0285 & 0.0364 & {\ul 0.0697} & \textbf{0.0751} & +7.76\% \\
 & \multicolumn{1}{l}{NDCG@10} & 0.0177 & 0.0193 & 0.0112 & 0.0240 & 0.0165 & 0.0198 & 0.0267 & 0.0139 & 0.0188 & {\ul 0.0409} & \textbf{0.0449} & +9.76\% \\
 & MRR & 0.0154 & 0.0170 & 0.0106 & 0.0207 & 0.0160 & 0.0181 & 0.0226 & 0.0122 & 0.0159 & {\ul 0.0368} & \textbf{0.0387} & +5.00\% \\ \midrule
 & Recall@5 & 0.0301 & 0.0266 & 0.0240 & 0.0551 & 0.0300 & 0.0411 & 0.0577 & 0.0243 & 0.0398 & {\ul 0.0703} & \textbf{0.0819} & +16.57\% \\
 & NDCG@5 & 0.0194 & 0.0173 & 0.0210 & 0.0377 & 0.0206 & 0.0298 & 0.0412 & 0.0153 & 0.0257 & {\ul 0.0473} & \textbf{0.0577} & +21.87\% \\
Toys & \multicolumn{1}{l}{Recall@10} & 0.0460 & 0.0447 & 0.0262 & 0.0797 & 0.0466 & 0.0658 & 0.0800 & 0.0393 & 0.0634 & {\ul 0.1055} & \textbf{0.1150} & +9.09\% \\
 & \multicolumn{1}{l}{NDCG@10} & 0.0245 & 0.0231 & 0.0231 & 0.0456 & 0.0260 & 0.0354 & 0.0481 & 0.0201 & 0.0332 & {\ul 0.0587} & \textbf{0.0684} & +16.53\% \\
 & MRR & 0.0216 & 0.0200 & 0.0221 & 0.0385 & 0.0244 & 0.0300 & 0.0415 & 0.0177 & 0.0280 & {\ul 0.0505} & \textbf{0.0584} & +15.55\% \\ \midrule
 & Recall@5 & 0.0214 & 0.0226 & 0.0302 & 0.0656 & 0.0485 & 0.0512 & 0.0677 & 0.0196 & 0.0306 & {\ul 0.0728} & \textbf{0.0811} & +11.48\% \\
 & NDCG@5 & 0.0144 & 0.0157 & 0.0186 & 0.0428 & 0.0309 & 0.0324 & 0.0461 & 0.0137 & 0.0205 & {\ul 0.0492} & \textbf{0.0553} & +12.36\% \\
Office & \multicolumn{1}{l}{Recall@10} & 0.0306 & 0.0338 & 0.0550 & 0.0989 & 0.0848 & 0.0856 & 0.1021 & 0.0310 & 0.0487 & {\ul 0.1023} & \textbf{0.1238} & +20.91\% \\
 & \multicolumn{1}{l}{NDCG@10} & 0.0173 & 0.0194 & 0.0266 & 0.0534 & 0.0426 & 0.0432 & 0.0572 & 0.0173 & 0.0264 & {\ul 0.0588} & \textbf{0.0690} & +17.36\% \\
 & MRR & 0.0162 & 0.0181 & 0.0268 & 0.0457 & 0.0408 & 0.0411 & 0.0502 & 0.0162 & 0.0229 & {\ul 0.0520} & \textbf{0.0592} & +13.96\% \\ \bottomrule
\end{tabular}
\end{table*}

\subsection{Overall Comparisons~(RQ1)}
We compare the recommendation performances of all models in Table~\ref{tab:overall_perf} and quantitatively demonstrate the superiority of \modelname. We obtain the following observations:
\begin{itemize}[leftmargin=*]
    \item \modelname achieves the best performance against all baselines in all metrics, demonstrating superior recommendation performance over existing methods. The relative improvements range from 3.56\% to 21.87\% in all metrics. We can also observe that improvements are consistent in MRR for measuring the entire recommendation list, ranging from 5.00\% to 13.96\%. We attribute improvements to several factors of \modelname: (1). the proposed multi-relational self-attention module provides additional related item pairs in pair-wise attentions calculation; (2). the regularization of intra-sequence item relationships modeling enhances and regularizes the item embeddings for SR; (3). the explorative related item pairs modeling enriches item embedding learning from training item-item transitions in sequences.
    \item Among three groups of methods, the sequential methods~(MoHR and \modelname) with item relationships modeling are the best. The sequential methods~(STOSA, SASRec, BERT4Rec, and Caser) perform the second best while static methods achieve the worst performance. This observation reveals that the temporal order plays a crucial role in the recommendation. It also uncovers that auxiliary item relationships can significantly benefit the recommendation.
    \item Among all static methods, graph-based methods~(KGIN and LightGCN) achieve the best performance in all datasets, which indicates the necessity of higher-order connected collaborative neighbors for users and items learning.
    \item From the comparison among all sequential baselines, we can observe that STOSA performs the best and then the SASRec. This observation verifies the effectiveness of the Transformer architecture. 
\end{itemize}

\subsection{Parameters Sensitivity~(RQ2)}
\begin{figure}
\begin{subfigure}[t]{0.241\textwidth}
    \
    \includegraphics[width=0.98\textwidth]{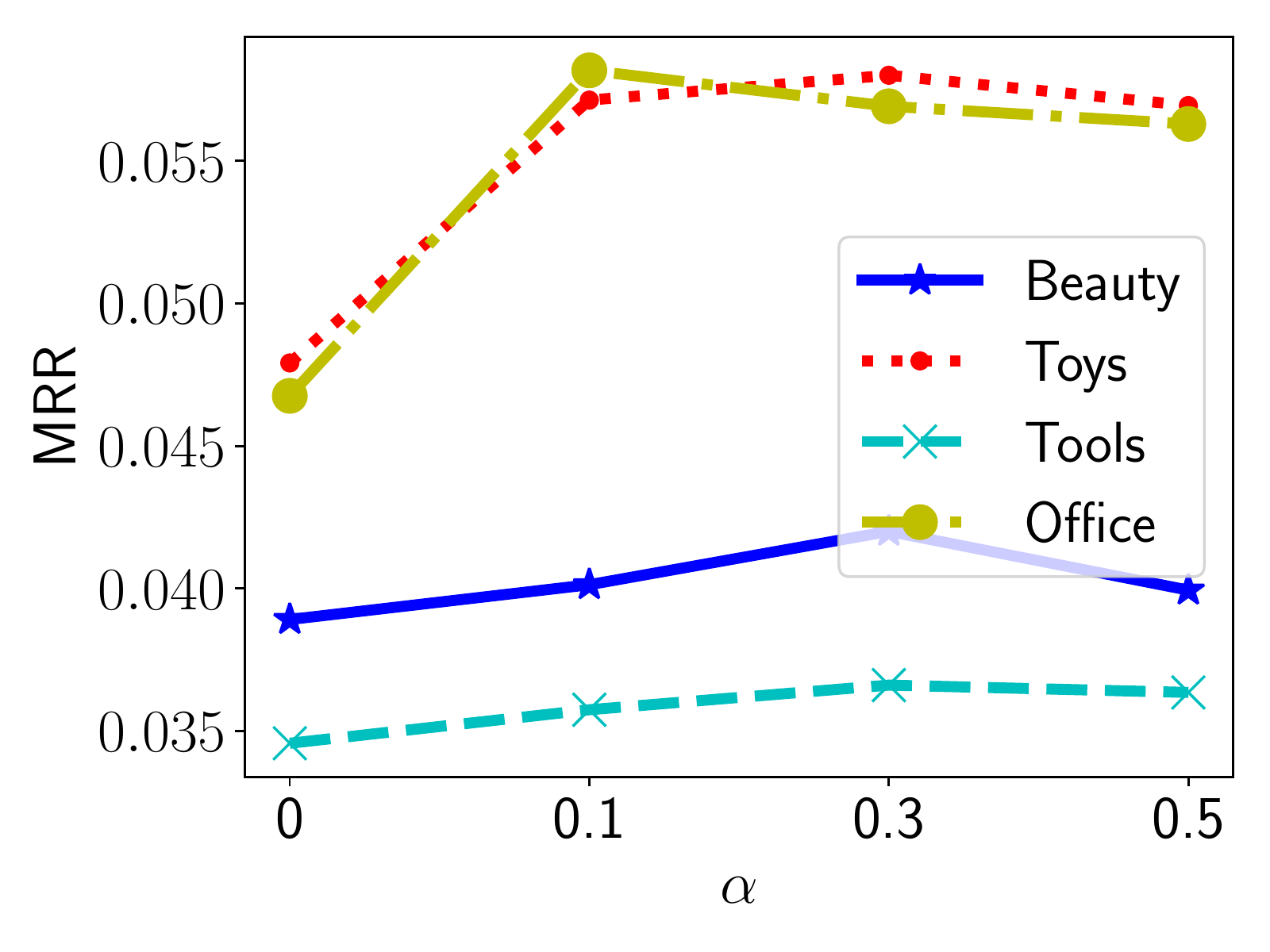}
    \caption{$\alpha$}
    \label{fig:mrr_over_alpha}
\end{subfigure}
\begin{subfigure}[t]{.241\textwidth}
    \includegraphics[width=0.98\textwidth]{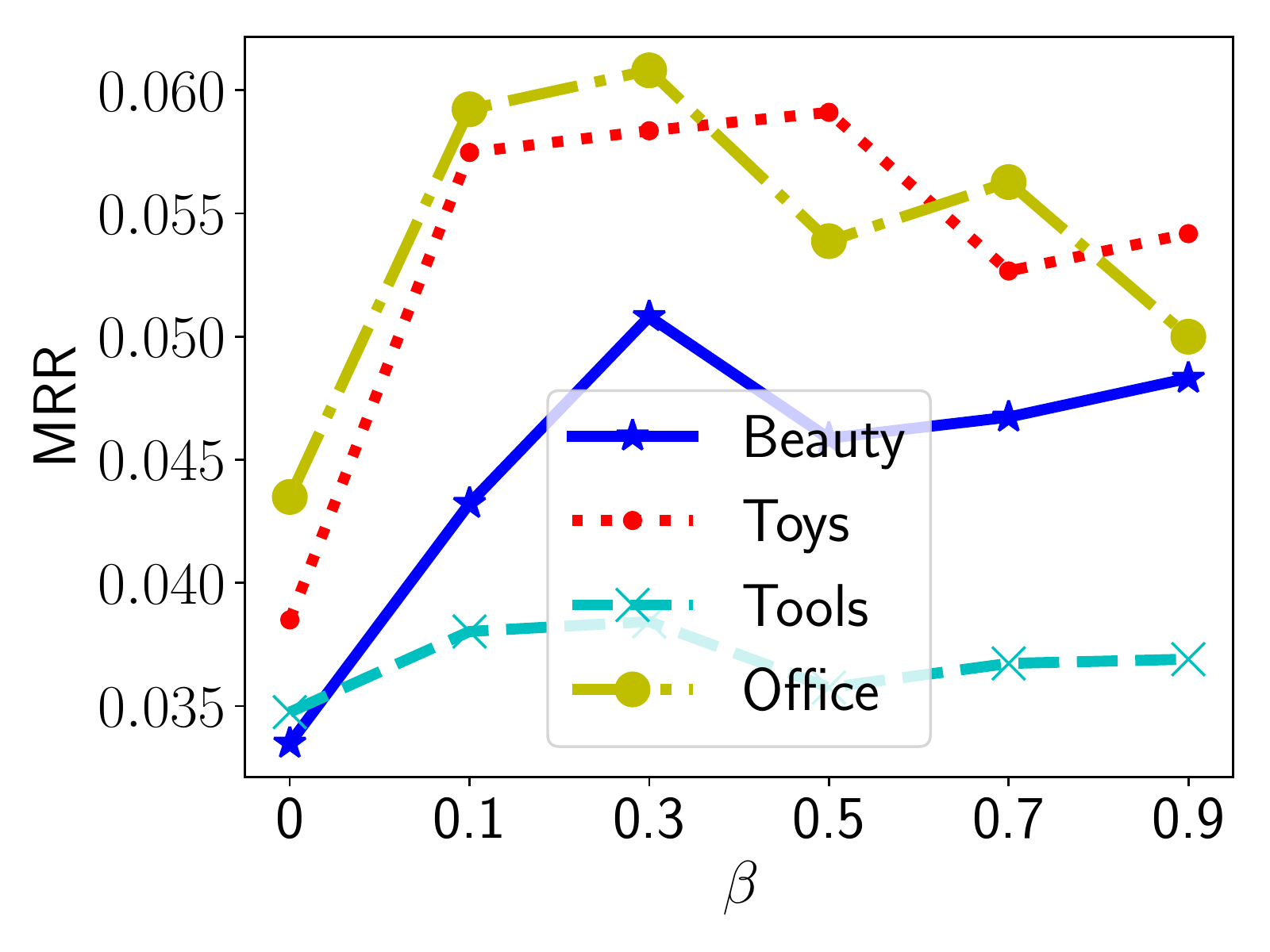}
    \caption{$\beta$}
    \label{fig:mrr_over_beta}
\end{subfigure}
\caption{MRR over different values of the weight $\alpha$ for $\mathcal{L}_{\text{intra}}$ and weight $\beta$ for $\mathcal{L}_{\text{inter}}$.}
\label{fig:mrr_over_param}
\end{figure}

In this section, we investigate the parameters sensitivity of $\alpha$ and $\beta$, which are weights for intra-sequence item relationships modeling regularization $\mathcal{L}_{intra}$ and inter-sequence related item pairs modeling regularization term $\mathcal{L}_{inter}$, respectively. The trend of $\alpha$ can be found in Figure~\ref{fig:mrr_over_alpha}. Figure~\ref{fig:mrr_over_beta} shows the sensitivity of $\beta$. Note that the special cases correspond to ablation studies of removing $\mathcal{L}_{intra}$ or $\mathcal{L}_{inter}$, when $\alpha=0$ or $\beta=0$, respectively.

Regarding the trend of $\alpha$, we can observe that the performance first increases and then drops as the value of $\alpha$ grows. 
We can see that the performance drops significantly when $\alpha=0$, from which we can conclude that the $\mathcal{L}_{intra}$ is crucial for improving recommendation. Moreover, increasing the weight of $\mathcal{L}_{intra}$ might still hurt the performance. The potential reason may be the sparsity of intra-sequence item relationships, in which some negative pairs might not be ground truth negatives but just unobserved positive item pairs.

For the sensitivity of $\beta$, we can observe that the performance first increases, and then the performance decreases as the $\beta$ grows. Compared with the special case where $\beta=0$, all nonzeros $\beta$ perform better. This observation verifies the necessity of inter-sequence related item pairs modeling in the optimization. Unlike $\alpha$, performances of $\beta$ have more fluctuations. The underlying reason is that the intra-sequence regularization enhances overlapping item pairs with item-item transitions. The larger weight of $\alpha$ increasingly fits only existing item-item transitions in sequences. However, $\mathcal{L}_{inter}$ consists of inter-sequences pairs, which cannot be found in any sequence. Moreover, additional related item pairs might introduce noises to item embedding learning.


\subsection{Ablation Study~(RQ3)}
We investigate the effectiveness of each proposed component in Table~\ref{tab:ablation_study}, including the intra-sequence regularization $\mathcal{L}_{\text{intra}}$ and inter-sequence regularization $\mathcal{L}_{\text{inter}}$. We demonstrate the necessity of these two components by observing the ranking performance after the removal of them. 

We remove each module from top to bottom and report the MRR performance. The followings are our observations:
\begin{itemize}[leftmargin=*]
    \item Removing either $\mathcal{L}_{\text{intra}}$ or $\mathcal{L}_{\text{inter}}$ reduces the recommendation performance. This verifies the necessity of these two components and their item relationships modeling capability.
    \item Removing inter-sequence module $\mathcal{L}_{\text{inter}}$ brings more negative impacts than the removal of $\mathcal{L}_{\text{intra}}$. This verifies our previous analysis in Table~\ref{tab:seq_rel_ratio} that the item-item transitions within sequences have limited overlapping with related item pairs. This scarce overlapping indicates the limited additional knowledge from intra-sequence. 
    \item The performance of removing both $\mathcal{L}_{\text{intra}}$ and $\mathcal{L}_{\text{inter}}$ is worse than SASRec. The reasons are twofold. First, there is no supervised signal to guide the multi-relational item attention calculation without $\mathcal{L}_{\text{intra}}$. The auxiliary item attention values are not optimized and poorly fit with the item transitions. Moreover, the absence of $\mathcal{L}_{\text{inter}}$ limits the model from capturing users' preferences from only the item transitions. 
\end{itemize}

\begin{table}[]
\centering
\caption{MRR of removing different modules in \modelname.}
\label{tab:ablation_study}
\begin{tabular}{@{}lllll@{}}
\toprule
Module & Beauty & Tools & Toys & Office \\ \midrule
\modelname & 0.0408 & 0.0387 & 0.0584 & 0.0592 \\
\modelname-$\mathcal{L}_{\text{intra}}$ & 0.0389 & 0.0346 & 0.0479 & 0.0468 \\
\modelname-$\mathcal{L}_{\text{inter}}$ & 0.0335 & 0.0347 & 0.0385 & 0.0435 \\
\modelname-$\mathcal{L}_{\text{intra}}$-$\mathcal{L}_{\text{inter}}$ & 0.0266 & 0.0165 & 0.0310 & 0.0381 \\
\bottomrule
\end{tabular}
\end{table}

\subsection{Improvements Analysis~(RQ4)}
We analyze the origins of improvements of \modelname by investigating performance differences in groups of users and items, separated by the number of interactions. It demonstrates the effectiveness of \modelname on cold start users and items and the capability of modeling high-order related item pairs.

\subsubsection{Performance w.r.t Sequence Length}
\begin{figure}
\begin{subfigure}[t]{0.241\textwidth}
    \
    \includegraphics[width=0.98\textwidth]{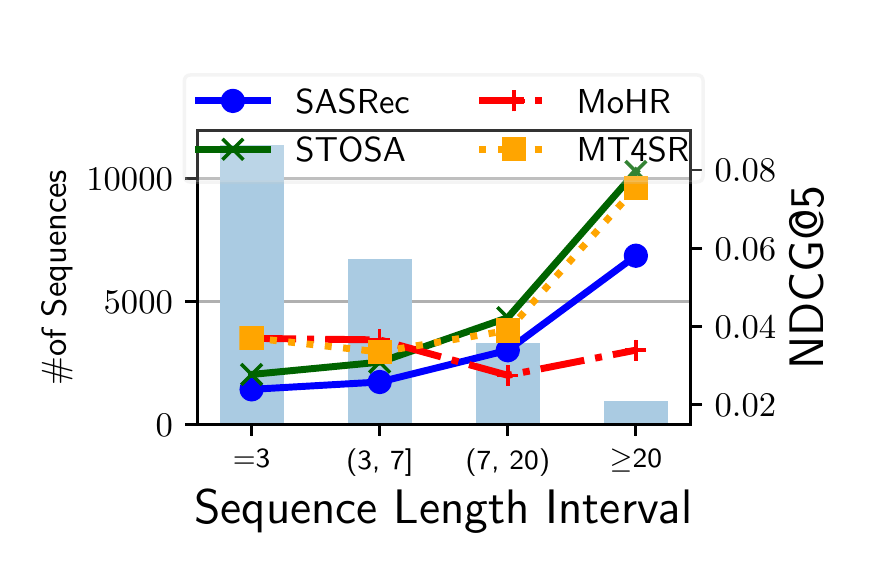}
    \caption{Beauty}
    \label{fig:beauty_item}
\end{subfigure}
\begin{subfigure}[t]{.241\textwidth}
    \includegraphics[width=0.98\textwidth]{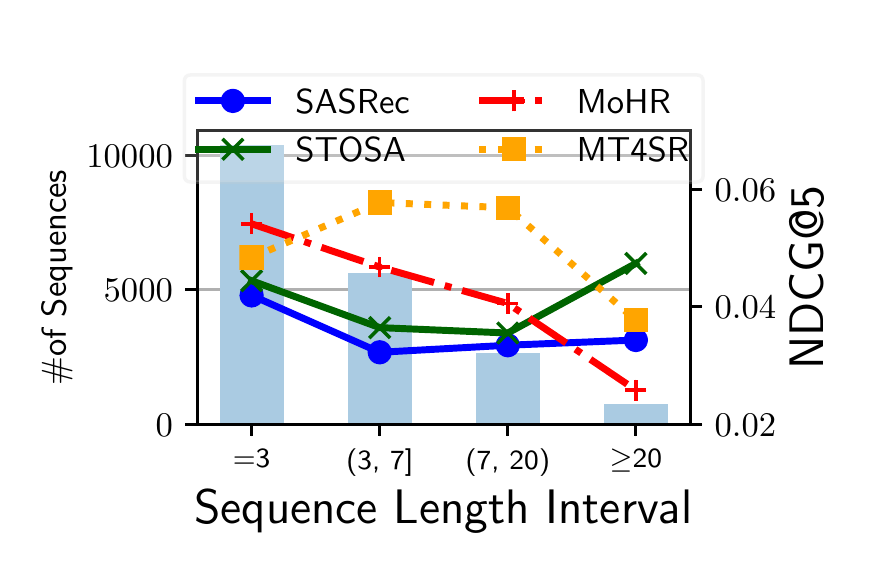}
    \caption{Toys}
    \label{fig:toys_item}
\end{subfigure}
\\
\begin{subfigure}[t]{0.241\textwidth}
    \
    \includegraphics[width=0.98\textwidth]{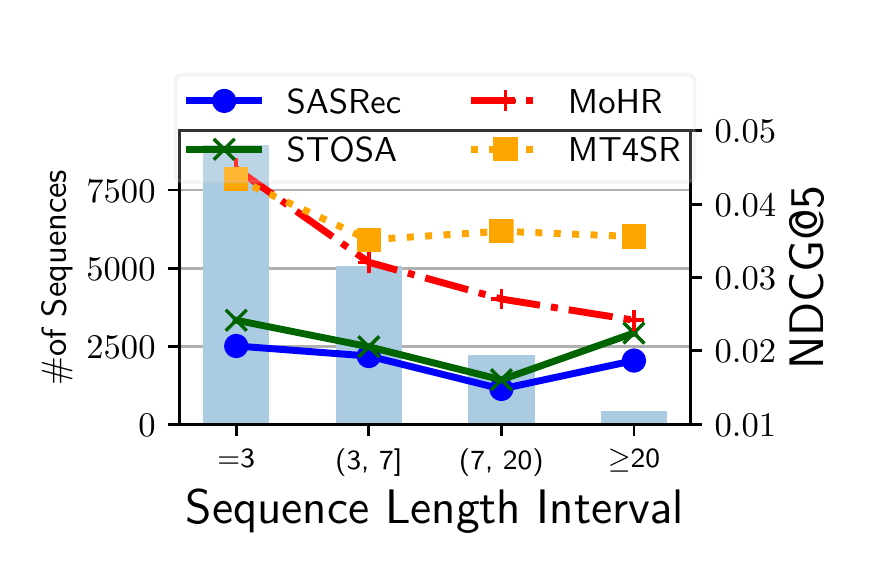}
    \caption{Tools}
    \label{fig:tools_item}
\end{subfigure}
\begin{subfigure}[t]{.241\textwidth}
    \includegraphics[width=0.98\textwidth]{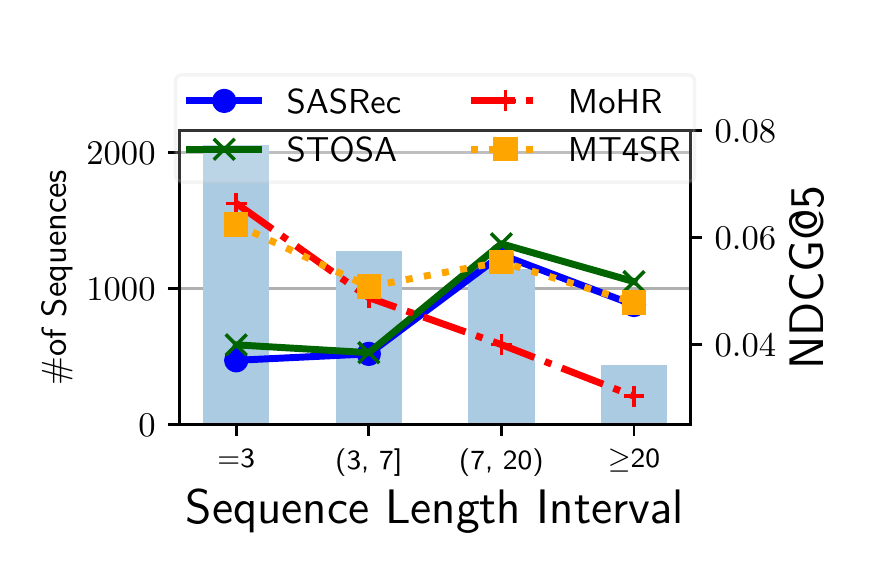}
    \caption{Office}
    \label{fig:office_item}
\end{subfigure}
\caption{NDCG@5 on different sequences based on length.}
\label{fig:ndcg5_seqlen}
\end{figure}

We separate users into sets based on the number of interactions in training, which is also sequence lengths of users. We report the average NDCG@5 on each group of users. Figure~\ref{fig:ndcg5_seqlen} shows sizes and the corresponding NDCG@5 of each group of users. The set with the shortest sequence length has most users, and sizes decrease as sequence lengths become longer. The models with auxiliary related item pairs~(\modelname and MoHR) significantly outperform STOSA and SASRec in short to medium lengths of sequences, with relative improvements of almost 200\%. However, \modelname and MoHR only achieve comparative performances in longest sequences. The reason is that auxiliary item relationships provide additional information for items.
This observation verifies the effectiveness of incorporating auxiliary item relationships on cold users. 
MoHR obtains the best performance in short sequences. However, its performance decreases drastically when sequence length becomes longer. The reason is MoHR only models first-order transitions. Unlike MoHR, \modelname learns both high-order item transitions and auxiliary item relationships.

\subsubsection{Performance w.r.t Item Popularity}
\begin{figure}
\begin{subfigure}[t]{0.241\textwidth}
    \
    \includegraphics[width=0.98\textwidth]{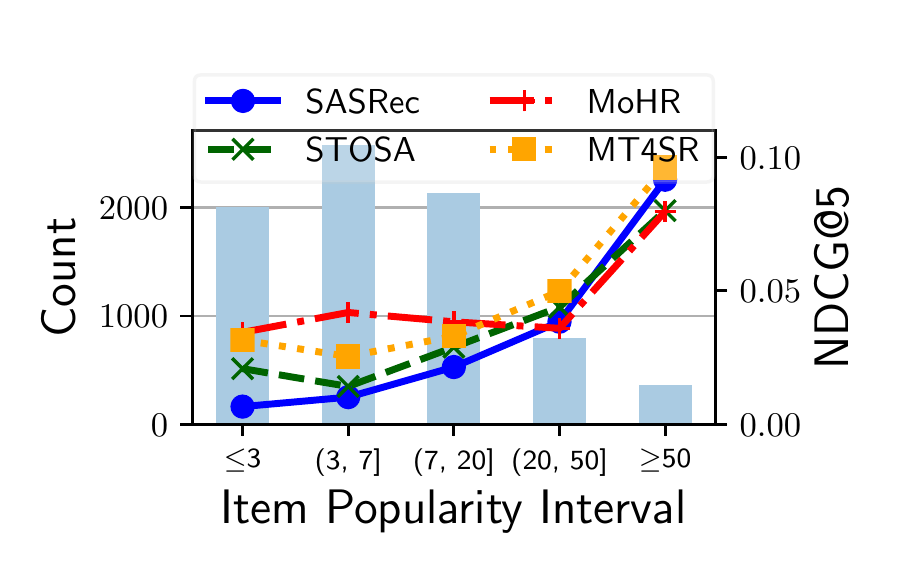}
    \caption{Beauty}
    \label{fig:beauty_item}
\end{subfigure}
\begin{subfigure}[t]{.241\textwidth}
    \includegraphics[width=0.98\textwidth]{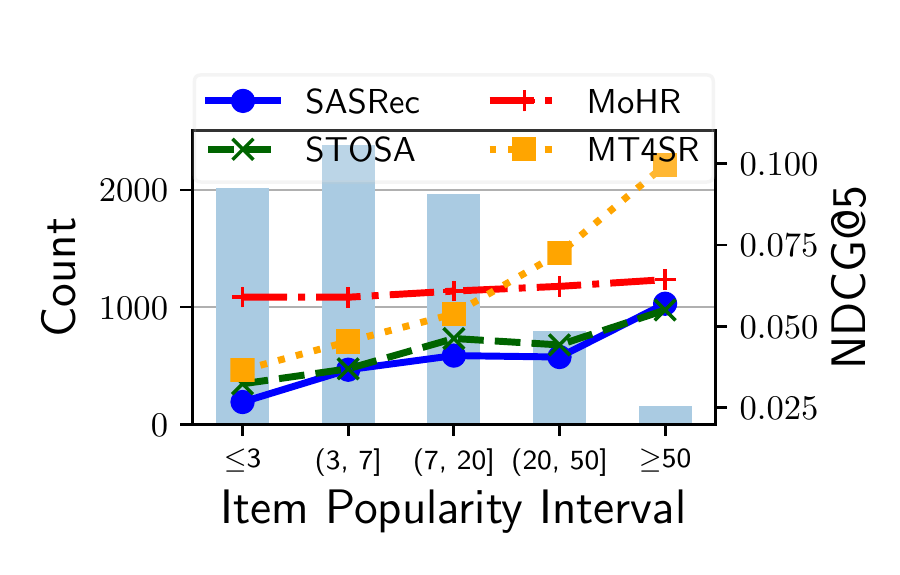}
    \caption{Toys}
    \label{fig:toys_item}
\end{subfigure}
\\
\begin{subfigure}[t]{0.241\textwidth}
    \
    \includegraphics[width=0.98\textwidth]{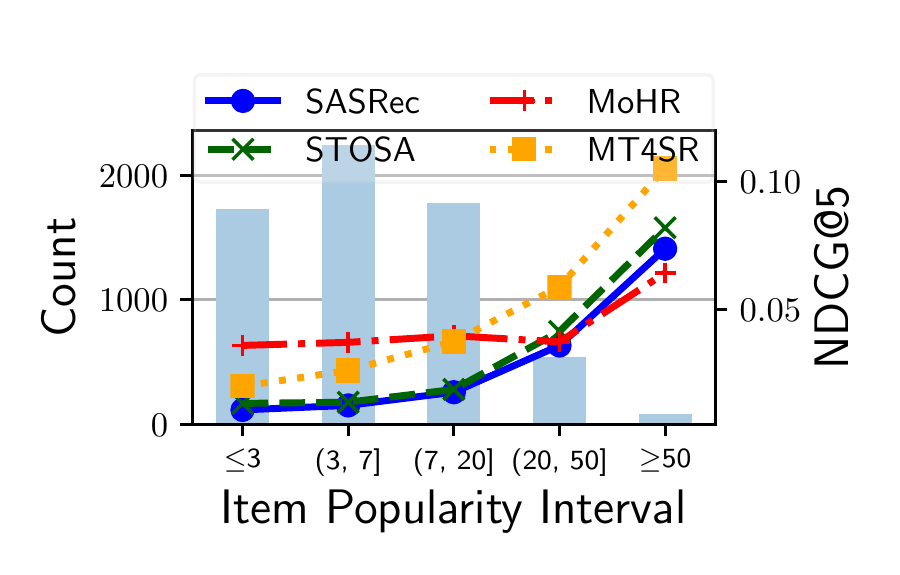}
    \caption{Tools}
    \label{fig:tools_item}
\end{subfigure}
\begin{subfigure}[t]{.241\textwidth}
    \includegraphics[width=0.98\textwidth]{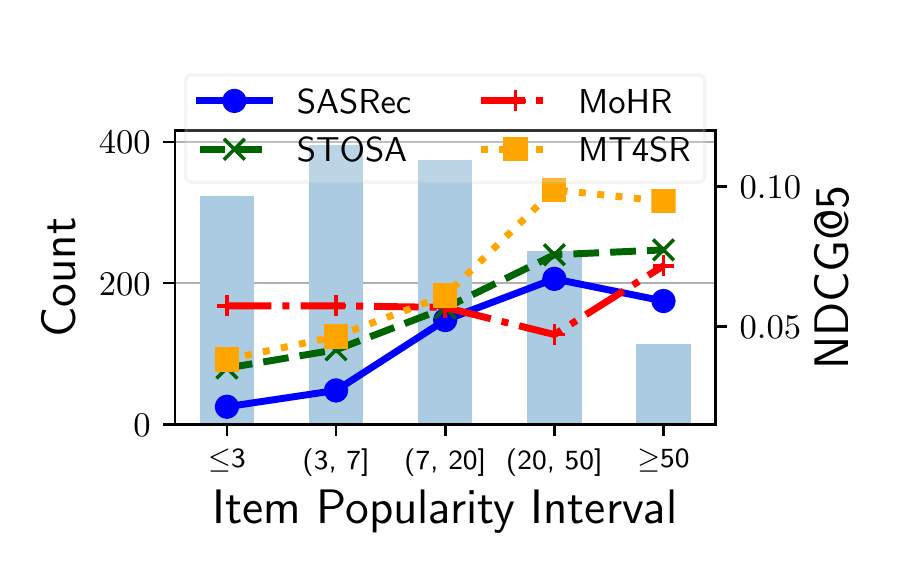}
    \caption{Office}
    \label{fig:office_item}
\end{subfigure}
\caption{NDCG@5 on different items based on popularity.}
\label{fig:item_pop}
\end{figure}
We separate items into groups based on popularity~(\textit{i.e.,} number of interacted users). We show the size and the average NDCG@5 in each group of items in Figure~\ref{fig:item_pop}. The size distributions are similar to those of users, where unpopular items are in the majority. 
Compared with models without auxiliary item relationships modeling~(STOSA and SASRec), \modelname and MoHR significantly improve performances on cold items. It demonstrates the effectiveness and necessity of incorporating item relationships. Unlike the sequence perspective, \modelname performs better on most popular items. 
Comparing \modelname and MoHR, \modelname performs significantly better than MoHR on most popular items. However, MoHR performs better than \modelname for cold items. The potential reasons for this observation are: (1). high-order transitions modeling of the multi-relational self-attention module connects more similar items for popular items; (2). related item pairs follow the power-law distributions, where popular items have most pairs.

\section{Conclusion}
This work proposes a novel and general Multi-relational Transformer \modelname for modeling high-order transitions and auxiliary item relationships simultaneously. To supervise the intra-sequence relatedness of item pairs, we also introduce a novel regularization measuring errors between related item pairs predictions and ground truth item pairs, guaranteeing accurate item relatedness self-attention calculations. We also explore inter-sequence item pairs with a novel regularization term. Extensive results and qualitative analysis on four real-world datasets demonstrate the effectiveness of \modelname and well support the superiority of \modelname in alleviating cold-start user and item issues and the capability of modeling high-order item relationships for SR.

\section*{Acknowledgment}
This paper was supported by the National Key R\&D Program of China through grant 2021YFB1714800, S\&T Program of Hebei through grant 20310101D, NSFC through grant 62002007, Natural Science Foundation of Beijing Municipality through grant 4222030, and the Fundamental Research Funds for the Central Universities. 
Philip S. Yu was supported by NSF under grants III-1763325, III-1909323, III-2106758, and SaTC-1930941.
For any correspondence, please refer to Hao Peng.


\bibliographystyle{ieeetr}
\bibliography{sample-base}

\begin{thebibliography}{10}

\bibitem{rendle2010factorizing}
S.~Rendle, C.~Freudenthaler, and L.~Schmidt-Thieme, ``Factorizing personalized
  markov chains for next-basket recommendation,'' in {\em Proceedings of the
  19th international conference on World wide web}, pp.~811--820, 2010.

\bibitem{ma2019hierarchical}
C.~Ma, P.~Kang, and X.~Liu, ``Hierarchical gating networks for sequential
  recommendation,'' in {\em Proceedings of the 25th ACM SIGKDD international
  conference on knowledge discovery \& data mining}, pp.~825--833, 2019.

\bibitem{vaswani2017attention}
A.~Vaswani, N.~Shazeer, N.~Parmar, J.~Uszkoreit, L.~Jones, A.~N. Gomez,
  {\L}.~Kaiser, and I.~Polosukhin, ``Attention is all you need,'' in {\em
  Advances in neural information processing systems}, pp.~5998--6008, 2017.

\bibitem{kang2018self}
W.-C. Kang and J.~McAuley, ``Self-attentive sequential recommendation,'' in
  {\em 2018 IEEE International Conference on Data Mining (ICDM)}, pp.~197--206,
  IEEE, 2018.

\bibitem{sun2019bert4rec}
F.~Sun, J.~Liu, J.~Wu, C.~Pei, X.~Lin, W.~Ou, and P.~Jiang, ``Bert4rec:
  Sequential recommendation with bidirectional encoder representations from
  transformer,'' in {\em Proceedings of the 28th ACM international conference
  on information and knowledge management}, pp.~1441--1450, 2019.

\bibitem{fan2022sequential}
Z.~Fan, Z.~Liu, Y.~Wang, A.~Wang, Z.~Nazari, L.~Zheng, H.~Peng, and P.~S. Yu,
  ``Sequential recommendation via stochastic self-attention,'' in {\em
  Proceedings of the ACM Web Conference 2022}, pp.~2036--2047, 2022.

\bibitem{kang2018recommendation}
W.-C. Kang, M.~Wan, and J.~McAuley, ``Recommendation through mixtures of
  heterogeneous item relationships,'' in {\em Proceedings of the 27th ACM
  International Conference on Information and Knowledge Management},
  pp.~1143--1152, 2018.

\bibitem{wang2021learning}
X.~Wang, T.~Huang, D.~Wang, Y.~Yuan, Z.~Liu, X.~He, and T.-S. Chua, ``Learning
  intents behind interactions with knowledge graph for recommendation,'' in
  {\em Proceedings of the Web Conference 2021}, pp.~878--887, 2021.

\bibitem{wang2019kgat}
X.~Wang, X.~He, Y.~Cao, M.~Liu, and T.-S. Chua, ``Kgat: Knowledge graph
  attention network for recommendation,'' in {\em Proceedings of the 25th ACM
  SIGKDD international conference on knowledge discovery \& data mining},
  pp.~950--958, 2019.

\bibitem{10.1145/3485447.3512273}
L.~Yang, Z.~Liu, Y.~Wang, C.~Wang, Z.~Fan, and P.~S. Yu, ``Large-scale
  personalized video game recommendation via social-aware contextualized graph
  neural network,'' in {\em Proceedings of the ACM Web Conference 2022},
  p.~3376–3386, 2022.

\bibitem{liu2021augmenting}
Z.~Liu, Z.~Fan, Y.~Wang, and P.~S. Yu, ``Augmenting sequential recommendation
  with pseudo-prior items via reversely pre-training transformer,'' in {\em
  Proceedings of the 44th international ACM SIGIR conference on Research and
  development in information retrieval}, pp.~1608--1612, 2021.

\bibitem{fan2021continuous}
Z.~Fan, Z.~Liu, J.~Zhang, Y.~Xiong, L.~Zheng, and P.~S. Yu, ``Continuous-time
  sequential recommendation with temporal graph collaborative transformer,'' in
  {\em Proceedings of the 30th ACM International Conference on Information \&
  Knowledge Management}, CIKM '21, p.~433–442, Association for Computing
  Machinery, 2021.

\bibitem{9750359}
C.~Wang, Y.~Liang, Z.~Liu, T.~Zhang, and P.~S. Yu, ``Pre-training graph neural
  network for cross domain recommendation,'' in {\em 2021 IEEE Third
  International Conference on Cognitive Machine Intelligence (CogMI)},
  pp.~140--145, 2021.

\bibitem{peng2020m2}
B.~Peng, Z.~Ren, S.~Parthasarathy, and X.~Ning, ``M2: Mixed models with
  preferences, popularities and transitions for next-basket recommendation,''
  {\em arXiv preprint arXiv:2004.01646}, 2020.

\bibitem{peng2022recursive}
B.~Peng, S.~Parthasarathy, and X.~Ning, ``Recursive attentive methods with
  reused item representations for sequential recommendation,'' {\em arXiv
  preprint arXiv:2209.07997}, 2022.

\bibitem{qiu2020exploiting}
R.~Qiu, Z.~Huang, J.~Li, and H.~Yin, ``Exploiting cross-session information for
  session-based recommendation with graph neural networks,'' {\em ACM
  Transactions on Information Systems (TOIS)}, vol.~38, no.~3, pp.~1--23, 2020.

\bibitem{he2020lightgcn}
X.~He, K.~Deng, X.~Wang, Y.~Li, Y.~Zhang, and M.~Wang, ``Lightgcn: Simplifying
  and powering graph convolution network for recommendation,'' in {\em
  Proceedings of the 43rd International ACM SIGIR conference on research and
  development in Information Retrieval}, pp.~639--648, 2020.

\bibitem{xin2019relational}
X.~Xin, X.~He, Y.~Zhang, Y.~Zhang, and J.~Jose, ``Relational collaborative
  filtering: Modeling multiple item relations for recommendation,'' in {\em
  Proceedings of the 42nd international ACM SIGIR conference on research and
  development in information retrieval}, pp.~125--134, 2019.

\bibitem{he2016fusing}
R.~He and J.~McAuley, ``Fusing similarity models with markov chains for sparse
  sequential recommendation,'' in {\em 2016 IEEE 16th International Conference
  on Data Mining (ICDM)}, pp.~191--200, IEEE, 2016.

\bibitem{quadrana2017personalizing}
M.~Quadrana, A.~Karatzoglou, B.~Hidasi, and P.~Cremonesi, ``Personalizing
  session-based recommendations with hierarchical recurrent neural networks,''
  in {\em Proceedings of the Eleventh ACM Conference on Recommender Systems},
  pp.~130--137, 2017.

\bibitem{zheng2019gated}
L.~Zheng, Z.~Fan, C.-T. Lu, J.~Zhang, and P.~S. Yu, ``Gated spectral units:
  Modeling co-evolving patterns for sequential recommendation,'' in {\em
  Proceedings of the 42nd International ACM SIGIR Conference on Research and
  Development in Information Retrieval}, pp.~1077--1080, 2019.

\bibitem{tang2018personalized}
J.~Tang and K.~Wang, ``Personalized top-n sequential recommendation via
  convolutional sequence embedding,'' in {\em Proceedings of the Eleventh ACM
  International Conference on Web Search and Data Mining}, pp.~565--573, 2018.

\bibitem{hidasi2015session}
B.~Hidasi, A.~Karatzoglou, L.~Baltrunas, and D.~Tikk, ``Session-based
  recommendations with recurrent neural networks,'' in {\em 4th International
  Conference on Learning Representations, {ICLR} 2016, San Juan, Puerto Rico,
  May 2-4, 2016, Conference Track Proceedings}, 2016.

\bibitem{devlin2018bert}
J.~Devlin, M.~Chang, K.~Lee, and K.~Toutanova, ``{BERT:} pre-training of deep
  bidirectional transformers for language understanding,'' in {\em Proceedings
  of the 2019 Conference of the North American Chapter of the Association for
  Computational Linguistics: Human Language Technologies, June 2-7, 2019,
  Volume 1}, pp.~4171--4186, Association for Computational Linguistics, 2019.

\bibitem{li2020time}
J.~Li, Y.~Wang, and J.~McAuley, ``Time interval aware self-attention for
  sequential recommendation,'' in {\em Proceedings of the 13th international
  conference on web search and data mining}, pp.~322--330, 2020.

\bibitem{lin2020fissa}
J.~Lin, W.~Pan, and Z.~Ming, ``Fissa: fusing item similarity models with
  self-attention networks for sequential recommendation,'' in {\em Fourteenth
  ACM Conference on Recommender Systems}, pp.~130--139, 2020.

\bibitem{fan2021modeling}
Z.~Fan, Z.~Liu, S.~Wang, L.~Zheng, and P.~S. Yu, ``Modeling sequences as
  distributions with uncertainty for sequential recommendation,'' in {\em
  Proceedings of the 30th ACM International Conference on Information \&
  Knowledge Management}, CIKM '21, (New York, NY, USA), p.~3019–3023,
  Association for Computing Machinery, 2021.

\bibitem{liu2020basket}
Z.~Liu, X.~Li, Z.~Fan, S.~Guo, K.~Achan, and S.~Y. Philip, ``Basket
  recommendation with multi-intent translation graph neural network,'' in {\em
  2020 IEEE International Conference on Big Data (Big Data)}, pp.~728--737,
  IEEE, 2020.

\bibitem{DBLP:journals/tois/PengZDYZY22}
H.~Peng, R.~Zhang, Y.~Dou, R.~Yang, J.~Zhang, and P.~S. Yu, ``Reinforced
  neighborhood selection guided multi-relational graph neural networks,'' {\em
  {ACM} Trans. Inf. Syst.}, vol.~40, no.~4, pp.~69:1--69:46, 2022.

\bibitem{wang2020make}
C.~Wang, M.~Zhang, W.~Ma, Y.~Liu, and S.~Ma, ``Make it a chorus: knowledge-and
  time-aware item modeling for sequential recommendation,'' in {\em Proceedings
  of the 43rd International ACM SIGIR Conference on Research and Development in
  Information Retrieval}, pp.~109--118, 2020.

\bibitem{zhao2021ugrec}
X.~Zhao, Z.~Cheng, L.~Zhu, J.~Zheng, and X.~Li, ``Ugrec: Modeling directed and
  undirected relations for recommendation,'' in {\em Proceedings of the 44th
  International ACM SIGIR Conference on Research and Development in Information
  Retrieval}, pp.~193--202, 2021.

\bibitem{huang2018improving}
J.~Huang, W.~X. Zhao, H.~Dou, J.-R. Wen, and E.~Y. Chang, ``Improving
  sequential recommendation with knowledge-enhanced memory networks,'' in {\em
  The 41st International ACM SIGIR Conference on Research \& Development in
  Information Retrieval}, pp.~505--514, 2018.

\bibitem{liu2022federated}
Z.~Liu, L.~Yang, Z.~Fan, H.~Peng, and P.~S. Yu, ``Federated social
  recommendation with graph neural network,'' {\em ACM Transactions on
  Intelligent Systems and Technology (TIST)}, vol.~13, no.~4, pp.~1--24, 2022.

\bibitem{wang2021dskreg}
Y.~Wang, Z.~Liu, Z.~Fan, L.~Sun, and P.~S. Yu, ``Dskreg: Differentiable
  sampling on knowledge graph for recommendation with relational gnn,'' in {\em
  Proceedings of the 30th ACM International Conference on Information \&
  Knowledge Management}, pp.~3513--3517, 2021.

\bibitem{yang2022large}
L.~Yang, Z.~Liu, Y.~Wang, C.~Wang, Z.~Fan, and P.~S. Yu, ``Large-scale
  personalized video game recommendation via social-aware contextualized graph
  neural network,'' in {\em Proceedings of the ACM Web Conference 2022},
  pp.~3376--3386, 2022.

\bibitem{bordes2013translating}
A.~Bordes, N.~Usunier, A.~Garcia-Duran, J.~Weston, and O.~Yakhnenko,
  ``Translating embeddings for modeling multi-relational data,'' {\em Advances
  in neural information processing systems}, vol.~26, 2013.

\bibitem{yang2014embedding}
B.~Yang, W.~Yih, X.~He, J.~Gao, and L.~Deng, ``Embedding entities and relations
  for learning and inference in knowledge bases,'' in {\em 3rd International
  Conference on Learning Representations, {ICLR} 2015, San Diego, CA, USA, May
  7-9, 2015, Conference Track Proceedings} (Y.~Bengio and Y.~LeCun, eds.),
  2015.

\bibitem{zhang2016collaborative}
F.~Zhang, N.~J. Yuan, D.~Lian, X.~Xie, and W.-Y. Ma, ``Collaborative knowledge
  base embedding for recommender systems,'' in {\em Proceedings of the 22nd ACM
  SIGKDD international conference on knowledge discovery and data mining},
  pp.~353--362, 2016.

\bibitem{wang2018ripplenet}
H.~Wang, F.~Zhang, J.~Wang, M.~Zhao, W.~Li, X.~Xie, and M.~Guo, ``Ripplenet:
  Propagating user preferences on the knowledge graph for recommender
  systems,'' in {\em Proceedings of the 27th ACM international conference on
  information and knowledge management}, pp.~417--426, 2018.

\bibitem{liu2017analogical}
H.~Liu, Y.~Wu, and Y.~Yang, ``Analogical inference for multi-relational
  embeddings,'' in {\em International conference on machine learning},
  pp.~2168--2178, PMLR, 2017.

\bibitem{ji2021survey}
S.~Ji, S.~Pan, E.~Cambria, P.~Marttinen, and S.~Y. Philip, ``A survey on
  knowledge graphs: Representation, acquisition, and applications,'' {\em IEEE
  Transactions on Neural Networks and Learning Systems}, 2021.

\bibitem{he2017translation}
R.~He, W.-C. Kang, and J.~McAuley, ``Translation-based recommendation,'' in
  {\em Proceedings of the eleventh ACM conference on recommender systems},
  pp.~161--169, 2017.

\bibitem{krichene2020sampled}
W.~Krichene and S.~Rendle, ``On sampled metrics for item recommendation,'' in
  {\em Proceedings of the 26th ACM SIGKDD International Conference on Knowledge
  Discovery \& Data Mining}, pp.~1748--1757, 2020.

\bibitem{rendle2012bpr}
S.~Rendle, C.~Freudenthaler, Z.~Gantner, and L.~Schmidt{-}Thieme, ``{BPR:}
  bayesian personalized ranking from implicit feedback,'' in {\em {UAI} 2009,
  Proceedings of the Twenty-Fifth Conference on Uncertainty in Artificial
  Intelligence, Montreal, QC, Canada, June 18-21, 2009} (J.~A. Bilmes and A.~Y.
  Ng, eds.), pp.~452--461, {AUAI} Press, 2009.

\end{thebibliography}

\end{document}